\documentclass[%
reprint,
groupedaddress,
 amsmath,amssymb,
 aps,
pra,
floatfix,
nofootinbib,
showkeys
]{revtex4-1}

\usepackage{graphicx}
\usepackage{dcolumn}
\usepackage{bm}
\usepackage{upgreek}

\begin{document}

\title{Quantum teleportation using highly coherent emission \\ from
telecom C-band quantum dots}

\author{M. Anderson$^{1,2}$}
\author{T. M\"{u}ller$^{1}$}
\email{tina.muller@crl.toshiba.co.uk}
\author{J. Huwer$^{1}$}
\author{J. Skiba-Szymanska$^{1}$}
\author{ A.B. Krysa$^{3}$}
\author{R.M. Stevenson$^{1}$}
\author{J. Heffernan$^{4}$}
\author{D.A. Ritchie$^{2}$}
\author{A.J. Shields$^{1}$}
\affiliation{$^{1}$ Toshiba Research Europe Limited, 208 Science Park, Milton Road, Cambridge, CB4 0GZ, UK}
\affiliation{$^{2}$ Cavendish Laboratory, University of Cambridge, JJ Thomson Avenue, Cambridge, CB3 0HE, UK}
\affiliation{$^{3}$ EPSRC National Epitaxy Facility, University of Sheffield, Sheffield, S1 3JD, UK}
\affiliation{$^{4}$ Departement of Electronic and Electrical Engineering, University of Sheffield, Sheffield, S1 3JD, UK}

\date{\today}

\begin{abstract}
A practical way to link separate nodes in quantum networks is to send photons over the standard telecom fibre network. This requires sub-Poissonian photon sources in the telecom wavelength band around 1550 nm, where the photon coherence time has to be sufficient to enable the many interference-based technologies at the heart of quantum networks. Here, we show that droplet epitaxy InAs/InP quantum dots emitting in the telecom C-band can provide photons with coherence times exceeding 1 ns even under non-resonant excitation, more than a factor two longer than values reported for shorter wavelength quantum dots under similar conditions. We demonstrate that these coherence times enable near-optimal interference with a C-band laser qubit, with visibilities only limited by the quantum dot multiphoton emission. Using entangled photons, we further show teleportation of such qubits in six different bases with average fidelity reaching 88.3$\pm$4\%. Beyond direct applications in long-distance quantum communication, the high degree of coherence in these quantum dots is promising for future spin based telecom quantum network applications.
\end{abstract}


\maketitle

\section{Introduction}

Quantum network technologies \cite{Kimble.2008}, ranging from quantum teleportation \cite{Fattal.2004} and its applications in quantum communication \cite{qrep.Briegel.1998,Nilsson.2013,Huwer.2017}, to heralded entanglement \cite{qinter.Bernien.2013} and quantum computing \cite{Knill.2001}, rely on interference of indistinguishable photons, and demand sources of highly coherent photons with sub-Poissonian statistics. To establish larger scale quantum networks over the standard telecom fibre network, the photons should further have a wavelength in the minimum loss telecom C-band around 1550~nm.\\
\indent Solid-state quantum systems, most prominently semiconductor quantum dots (QDs) and defects in diamond, are well established candidates for emerging quantum technologies. Significant work in this field has developed the basic building blocks of a quantum network, such as near-ideal photon sources \cite{somaschi.2016}, deterministic entanglement between distant spins \cite{humphreys.2018}, and teleportation of qubits \cite{Nilsson.2013}. However, these systems have in common that their wavelength in the visible and near infrared up to about 900 nm prevents wider integration using the standard optical fibre infrastructure due to the strong attenuation ($>$ 1dB/km) at those wavelengths. Recent efforts have pushed QD emission to the telecom O-band at 1310 nm \cite{Huwer.2017} (0.3 dB/km), and the C-band (0.15 dB/km), where single and entangled photon pair sources were reported \cite{kim.2005,olbrich.2017,Muller.2018}. However, the coherent emission necessary for all interference-based quantum network applications has not yet been reported from any solid-state single photon emitter in that band.\\
\indent Alternatively, technologies based on wavelength conversion of single photon sources or non-linear optical processes have successfully been used to reach the telecom C-band and demonstrate two-photon interference \cite{weber.2018} as well as teleportation over several kilometers \cite{sun.2016,valivarthi.2016}, respectively. But these approaches are limited in efficiency due to the losses during the conversion process, or the Poissonian statistics underlying attenuated laser pulses that therefore require an additional layer of complexity to guarantee optimal operation.\\
\indent Here, we demonstrate coherent emission from InAs/InP QDs emitting in the standard telecom window, through measurement of single and two-photon interference. We show further the power and utility that this photonic platform offers in quantum information by teleporting a telecom C-band polarization encoded qubit.\\

\section{Singe-photon interference}
\begin{figure*}
\centering
\includegraphics[width=0.6\textwidth]{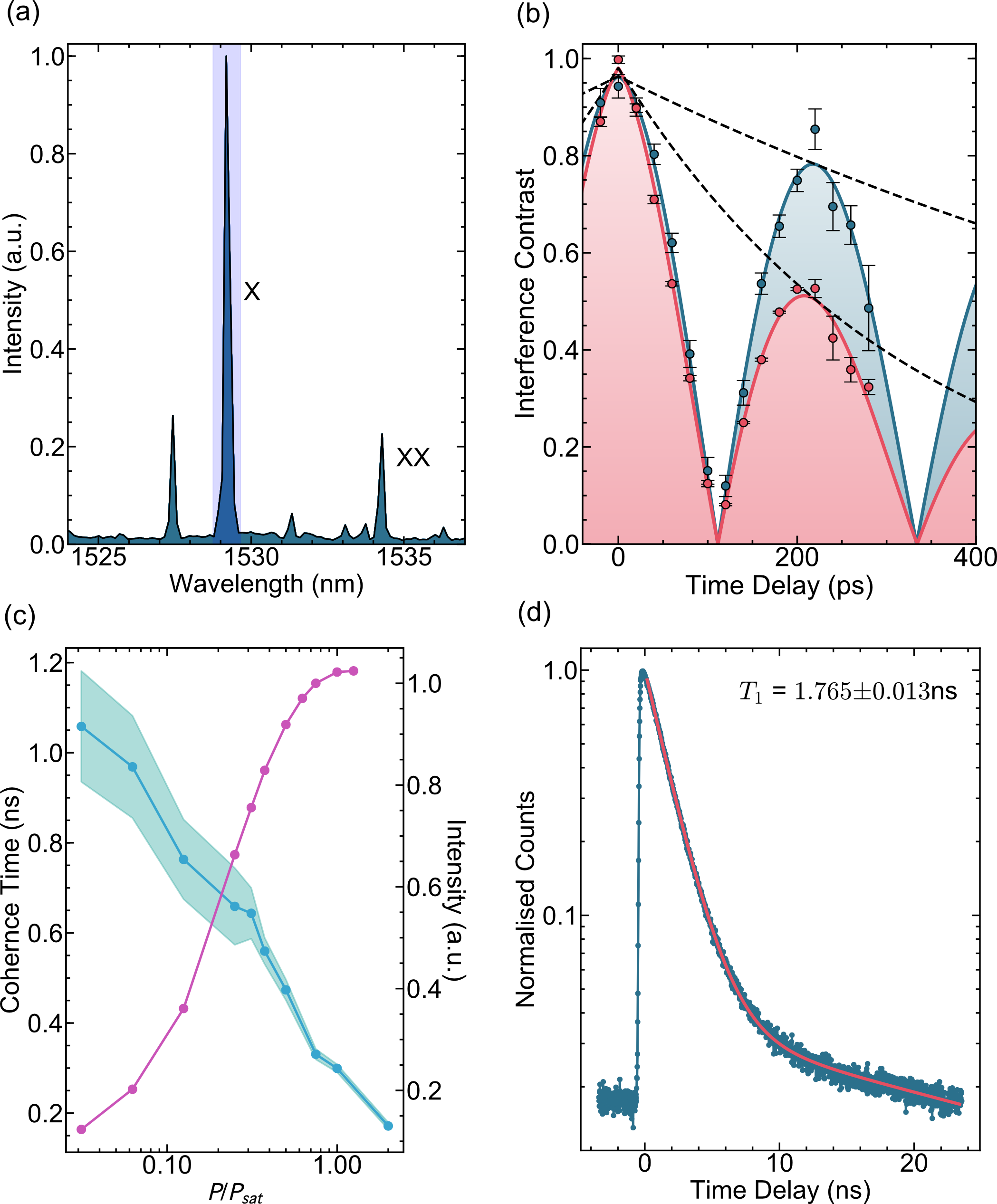} 
\label{fig:singlephoton}
\caption{Single-photon interference. (a) Spectrum of the quantum dot measured at $P_{sat}$ with exciton ($X$) and biexciton ($XX$) labelled. The shaded area denotes the part of the spectrum used for these measurements. (b) Single-photon interference measurements of the $X$ at 0.75$P_{sat}$ (pink) and 0.03$P_{sat}$ (blue). The visibility oscillates at the beat frequency corresponding to the FSS of the $X$ transition. Solid lines are fits to the data according to a Fourier transform of a double Lorentzian. (c) Coherence time ($T_{2}$) extracted from the fits (blue) and integrated intensity (pink) as a function of excitation power. Shaded areas denote standard error. (d) Time-resolved intensity of $X$. The solid line is a fit to a double exponential decay.}
\end{figure*}
\indent The experiments we report here use metalorganic vapor-phase epitaxy grown droplet InAs QDs on an InP substrate. To increase photon extraction efficiency, the QDs are embedded in a planar cavity between p-doped repeated layers of distributed Bragg reflectors on the top and n-doped layers at the bottom. The sample is similar to that used in our recent work \cite{Muller.2018}. The device is operated at 8.5 K in a temperature controlled Helium vapor cryostat. Emission from a single QD is collected using a standard confocal microscope (NA=0.68) coupled to a single-mode fibre. We excite the QD using a non-resonant continuous-wave laser diode at 785 nm.
A photoluminescence spectrum taken at the pump power where the neutral exciton ($X$) intensity saturates ($P_{sat}$) is shown in Fig.~1(a). We observe several isolated emission lines, and identify the $X$ and biexciton ($XX$) transitions  by polarization spectroscopy and intensity correlations.

\indent The coherence time of single photons can be measured with high temporal resolution through first-order field-correlation measurements using Fourier Transform Spectroscopy \cite{Zwiller.2004}. For this we use a single-mode optical fibre-based Michelson interferometer, where an optical delay line allows us to vary the delay between the arms coarsely and a fibre stretcher to measure interference fringes. When one arm of the interferometer is varied using the fibre stretcher, we measure single-photon interference as oscillations in the output intensity. Along with the delay line and fibre stretcher, Faraday mirrors are used at the end of both arms to compensate for polarization rotations and fluctuations in the fibres. The interference fringe visibility is defined as
\begin{equation}\label{spi}
V=\frac{I_{max}-I_{min}}{I_{max}+I_{min}},
\end{equation}
where $I_{max}$ and $I_{min}$ correspond to the cases of constructive and destructive interference, respectively. Interference fringes are then fitted to extract this visibility at different coarse delays.\\
\indent Two example visibility measurements performed on the quantum dot $X$ transition at excitation powers of 0.75$P_{sat}$ and 0.03$P_{sat}$ are given in Fig. 1(b), where the latter is chosen as the lowest power with a sufficient signal-to-noise ratio for the measurement. For both powers, the decay of the visibility displays a characteristic beating pattern in the fringe contrast resulting from the fine-structure splitting (FSS) of the $X$. However, the envelope of the oscillations decays more slowly for lower driving powers. The Fourier transform of a double Lorentzian was used to fit our data, and has the form
\begin{equation}\label{contrast}
A_{0}\exp(-|\Delta\tau|/T_{2})\cos(\Delta E\Delta\tau/\hbar)
\end{equation}
where $A_{0}$ is the fringe contrast at zero delay, $T_2$ is the coherence time,  and $\Delta E$ is the central energy difference. The fringe contrast $A_{0}$ accounts for experimental imperfections, but remains close to unity for all measurements presented here. From the fit we extract a FSS of $\Delta E=18.5~\pm~0.2$ $\upmu$eV and a coherence time  $T_{2}=331~\pm~12$ ps at 0.75$P_{sat}$, where the uncertainty is given by the standard error extracted from a least-squares fitting routine.  At the lower pump power, the coherence time remarkably increases to $T_{2}=1058~\pm~134$ ps, a factor of two longer than the highest previously reported values for quantum dot emission at any wavelength under this type of excitation scheme \cite{solomon.2010,Patel.2010}, as well as exceeding those measured using resonant p-shell excitation \cite{Gold.2014}, resonant two-photon excitation \cite{muller2014demand} and approaching those measured using strict resonant s-shell excitation \cite{Moody.2016}. \\
\indent The power dependence of $T_{2}$ is shown in Fig. 1(c), where coherence times are longer and dephasing is reduced for lower pump powers due to a more stable charge environment in the vicinity of the QD under these conditions \cite{Berthelot.2006}. Increasing the excitation power leads to brighter emission from the quantum dot (Fig. 1(c)) at the cost of increased charge fluctuations and spectral diffusion. We note however that even at maximum $X$ brightness at $P_{sat}$, the measured coherence time ($T_{2}=299~\pm~9$ ps) is still sufficient for applications such as a quantum relay \cite{Huwer.2017}. The fact that our QD maintains a high level of temporal coherence even under saturated pumping conditions where the intensity is the brightest for the $X$ is surprising, especially with the given excitation scheme where one would typically expect to observe a relatively high level of dephasing. We hypothesise that the reduced dephasing in our QDs is due to the two main differences to previous QD systems: the droplet epitaxy growth mode leads to a reduced wetting layer and fewer abundant charges compared to the more established Stransky-Krastanov growth \cite{joanna.FSS}, while the lower electron mobility and diffusion coefficient of the InP matrix compared to the traditional GaAs might further contribute to a more stable charge environment.\\
\indent A Fourier transform-limited photon exhibits a coherence time $T_{2}=2T_{1}$, where $T_1$ is the radiative lifetime of the transition. In this limit, the emitter is fundamentally free from external dephasing processes. A measurement of the exciton radiative lifetime, shown in Fig. 1(d), determines $T_{1}=1.765~\pm~0.013$ ns and thus $T_{2}/2T_{1} = 0.298~\pm~0.038$. Notably, this is comparable to values obtained in shorter wavelength systems under non-resonant as well as quasi-resonant p-shell excitation \cite{Gold.2014}, which is less disruptive than the non-resonant excitation scheme used here. Only under strict resonant s-shell excitation has this value been shown to approach the Fourier limit \cite{Moody.2016}. The radiative lifetime $T_{1}$ in our case could further be reduced by enhancing the cavity coupling of the quantum dot through the Purcell effect \cite{purcell.qd}, bringing the emission even closer to the transform limit \cite{cavity.enhanced}.

\section{Two-photon interference with a laser}
\begin{figure*}
\includegraphics[width=0.75\textwidth]{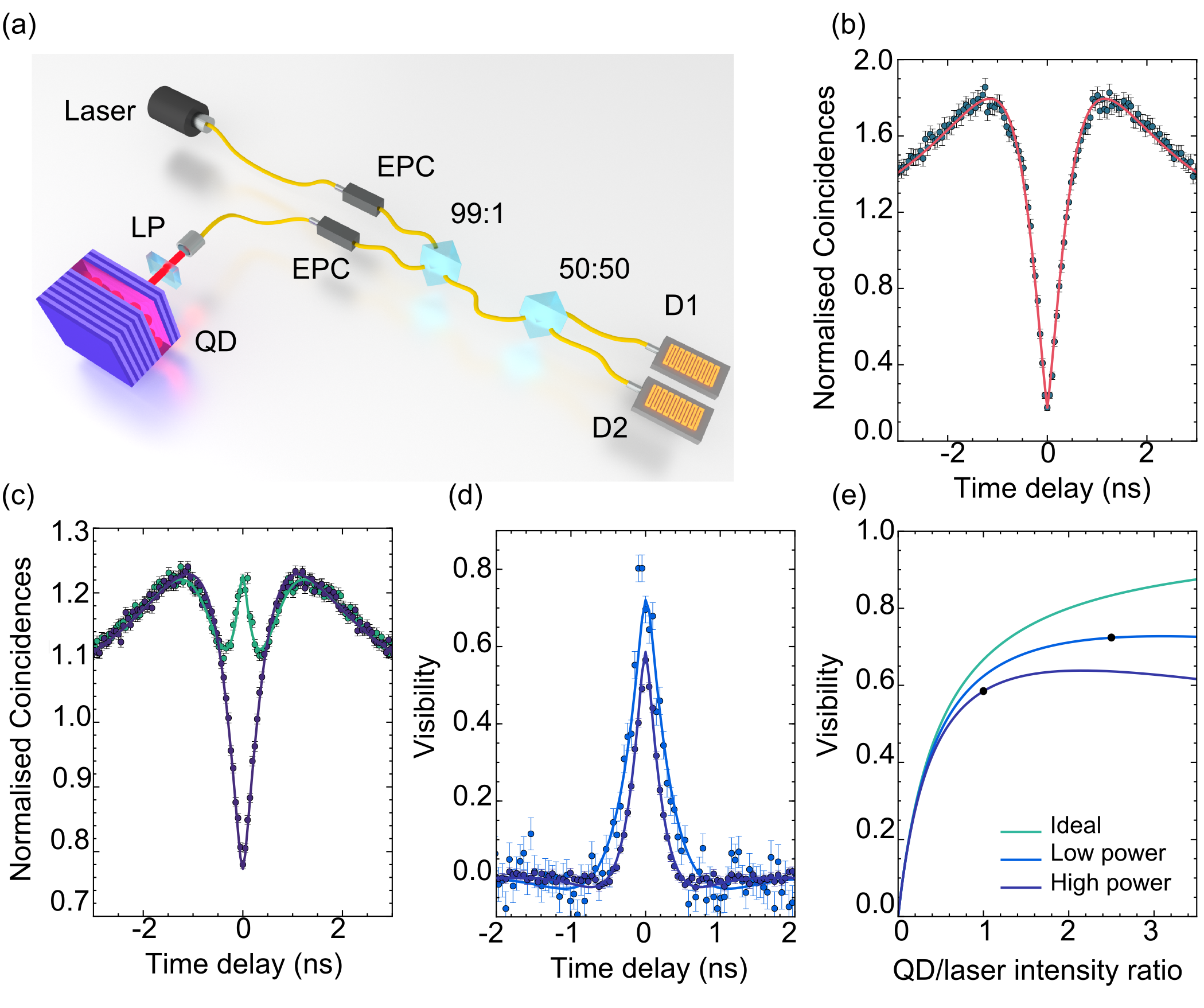} 
\label{fig:tpi}
\caption{Two-photon interference of a laser qubit.(a) Experimental schematic. A laser photon is interfered with a QD photon on a 99:1 unbalanced beamsplitter. Correlation measurements are performed on photons exciting through the same port of the beamsplitter using a standard Hanbury-Brown and Twiss (HBT) with a 50:50 beamsplitter and superconducting nanowire single photon detectors (D1 \& D2). The photon polarization is controlled using a linear polariser (LP) and electronic polarization controllers (EPCs). (b) Autocorrelation of the QD (black dots) fitted to a 4-level model (red solid line). (c) Correlation measurements of TPI for indistinguishable (green) and distinguishable (purple) photons. (d) Corresponding visibility of the TPI for the high (purple) and low (blue) intensity regimes discussed in the text. (e) Calculated TPI visibility as a function of the relative QD/laser intensity ratio $\eta/\alpha^{2}$  for the ideal case where the multi-photon emission probability $g^{(2)}(0)=0$ (green), and for the multi-photon emission probabilities $g^{(2)}(0) = 0.177$ (purple), and $0.095$ (blue) corresponding to recorded values at $P_{sat}$  and  at $P_{sat}/20$, respectively. Black points denote experimental values extracted from the fit in (d). There is no correction for detection response.}
\end{figure*}
The most fundamental quantum network technology enabled by the long coherence times is the interference of two photons. To demonstrate this, we interfere a quantum dot with a telecom-wavelength laser qubit. This is the first step in the quantum teleportation protocol presented in Section IV, and we pick a quantum dot for this experiment that will be suitable teleportation as well. In addition to the long temporal coherence time, we therefore require a dot with low FSS, leading to high entanglement fidelity, and also high brightness to collect the necessary correlation statistics. We typically find coherence times exceeding 150 ps in 80$\%$ of dots in our sample, even when driving at $P_{sat}$ to extract maximum count rates. Thus, we focus on choosing at dot with sufficient FSS. For the dot used in the remainder of the paper, we measured a FSS of $5.7~\pm~0.2$ $\upmu$eV, a maximum entanglement fidelity of $91.0~\pm~0.5$\%  and saturation count rates typically around 300k counts/s for $X$. \\
A schematic of the two-photon interference (TPI) measurement is shown in Fig. 2(a). Photons which are identical in all degrees of freedom will coalesce and leave through one port of a beamsplitter \cite{Hong.1987}, after which correlation measurements are performed. If the laser and dot photons are distinguishable (due to their orthogonal polarization), we expect a dip at zero delay due to the single photon nature  of the photon from the dot. If however the input photons are indistinguishable (in the co-polarised case), we expect to see bunching at zero delay due to two-photon interference superimposed on the dip.  This can be described by a modified version of the two-photon correlation function for dissimilar photon sources presented by Bennett \textit{et al.} \cite{Bennett.2009}, taking into account our experimental setup and background contributions. The form of the correlation function is then
\begin{multline}\label{tpi}
g^{(2)}(\tau)=1 + \Big[2\eta\alpha^{2}\big(1+e^{-\vert \tau\vert/T_{2}}\cos(\Delta E_{L}\tau/\hbar)\cos^2\phi\big) \\+\eta^{2}\big(g^{(2)}_{QD}(\tau)-1\big)+\alpha^{4}\Big]\\\times(\eta +\alpha^{2}+\beta)^{-2}
\end{multline}
where $\eta$ is the dot intensity, $\alpha^{2}$ is the laser intensity, $\beta$ is the background contribution, $\Delta E_{L}$ is the detuning between QD and laser photon, and $\phi$ gives the polarization difference between the two photons ($\phi=0$ and $\phi=\pi/2$ for co- and cross-polarised photons, respectively). This model is convolved with a Gaussian response function with $125$ ps FWHM to account for the measured detector response. The interference visibility can be calculated from the contrast of the co- and cross-polarised correlation measurements as $V(\tau) = g^{(2)}_{\|}(\tau)/g^{(2)}_{\bot}(\tau) - 1$, where $g^{(2)}_{\bot}(\tau)$ and $g^{(2)}_{\|}(\tau)$  are the correlation functions for the co- and cross-polarised cases, respectively.\\
\indent We start by measuring the quantum dot autocorrelation $g_{QD}^{(2)}(\tau)$ for the dot chosen for this experiment, to confirm the single photon nature of the emission. The autocorrelation of the QD at saturation power is shown in Fig. 2(b) and is described for this power by a 4-level model \cite{Kitson.1998} as
\begin{multline}\label{gtwo}
g^{(2)}_{QD}(\tau) = 1 - (1+X+Y)e^{-\vert \tau\vert/\tau_{1}}\\+ Xe^{-\vert \tau\vert/\tau_{2}}+ Ye^{-\vert \tau\vert/\tau_{3}},
\end{multline}
where $X$ and $Y$ denote the coupling rates from the additional levels and $\tau_{1}$, $\tau_{2}$ and $\tau_{3}$ are the timescales of the respective decay processes. Note that for lower powers, $Y\rightarrow0$ and the behavior reduces to a standard 3-level model. A full background corrected function is then defined as
\begin{equation}\label{gtwoplusbg}
g^{(2)}_{HBT}(\tau) =  \frac{g^{(2)}_{QD}(\tau) + 2\beta +\beta^{2}}{(1 + \beta)^{2}},
\end{equation}
where $\beta$ is the contribution of the uncorrelated background intensity. Equation \ref{gtwoplusbg} is used to fit the autocorrelation data in Fig. 2(b) and extract the background contribution for the system. From the fit we extract the multi-photon emission probability $g^{(2)}(0) = 0.177$ at $P_{sat}$, and a similar measurement at  $P_{sat}/20$ gives $g^{(2)}(0) =0.095$  (data not shown).\\
\indent We perform TPI for realistic teleportation conditions, where the QD is excited at $P_{sat}$ and $\eta/\alpha^{2} = 1$, as well as for more ideal conditions, where the driving power is reduced to $P_{sat}/20$, and $\eta/\alpha^{2} = 2.5$. Here, we excite the QD using a below-band non-resonant laser at 1309nm, where we see an increase in the neutral transition intensity, to increase the $g^{(2)}(\tau)$ statistics. This does not affect the measured coherence times.
As an example, the resulting correlation measurements for the higher power parameters are shown in Fig. 2(c). The calculated visibility for both sets of parameters can be seen in Fig. 2(d), where the peak values of $58.6~\pm~0.5\%$ and $72.4~\pm~0.7\%$ are extracted from the fit (solid lines). Again, these results surpass some of those seen in similar experiments at shorter wavelength under quasi-resonant excitation \cite{Thoma.2016}, and is expected to increase only when moving to pulsed excitation with a pulse separation shorter than the timescale of the spectral diffusion. From the fits we can also further determine $\tau_{c}=294~\pm~9$ ps at $P_{sat}$ and $\tau_{c}=471~\pm~29$ ps at $P_{sat}/20$. The coherence time at $P_{sat}$ is in agreement with the value measured using the MI for this dot ($262\pm19$ ps, data not shown), and is also comparable to the value shown by the dot presented in Fig. 1 at this power.\\
\indent In practice, both the multi-photon probability of the QD $( g^{(2)}(0))$ and the QD/laser intensity ratio $\eta/\alpha^{2}$  limit the maximum achievable visibility, as can be seen from Equation \ref{tpi}. In our case, for both power settings, the visibility is above 85\% of the ideal value where $g^{(2)}(0) = 0$, and is well described when taking the finite multi-photon emission at the respective powers into account, as shown in Fig. 2(e).

\section{Quantum teleportation}
\begin{figure*}
\includegraphics[width=0.75\textwidth]{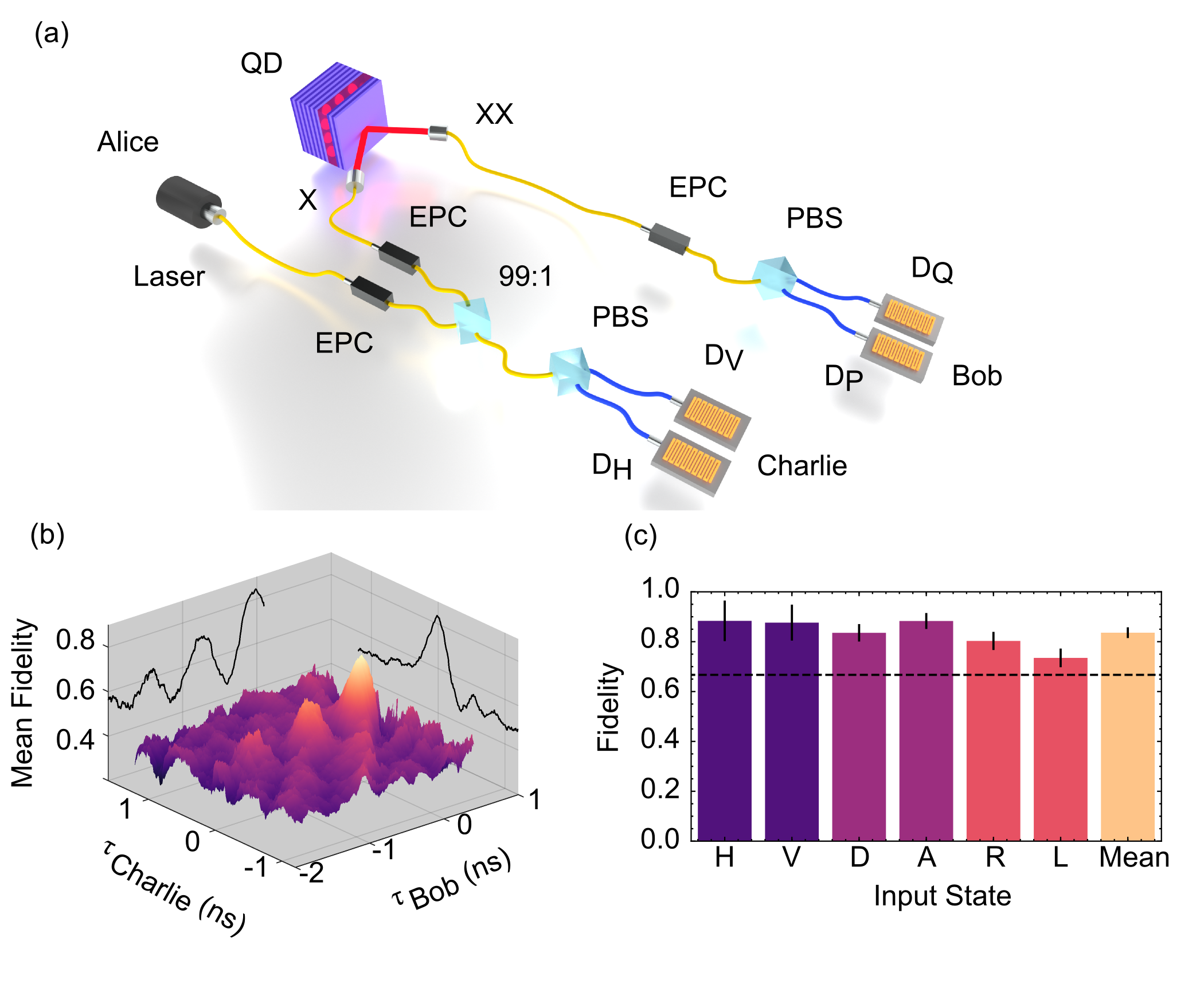} 
\label{fig:qteleport}
\caption{Quantum teleportation of a laser qubit. (a) Experimental schematic. An input polarization encoded laser qubit is interfered with an $X$ photon on a 99:1 beamsplitter. At Charlie, a Bell state measurement is performed using a polarising beamsplitter (PBS) and superconducting single-photon detectors ($D_{H}$ \& $D_{V}$). Bob then uses polarization resolved correlations to analyse the output state in different polarization bases at detectors ($D_{P}$ \& $D_{Q}$). (b) Mean fidelity coincidence map for the most significant post-selection window size. The peak fidelity at $\tau_{Charlie}=\tau_{Bob}=0$ is $83.6~\pm~2.2\%$. (c) Individual fidelities for the 6 input polarization states and the corresponding mean fidelity. Each basis is above the classical limit of $2/3$ (dotted line) for the 6-state teleportation protocol.}
\end{figure*}
One direct application of this type of interference is the teleportation of a qubit, a fundamental operation in quantum information technologies. The larger the temporal coherence of the interfering photons, the more photons are teleported with high fidelity.\\
\indent The quantum teleportation protocol relies on the distribution of an entangled pair of photons between an intermediate station, Charlie, and Bob. A Bell state measurement is then performed on the input qubit and one of the entangled photons (Charlie). This measurement projects the wavefunction of Bob's photon into the state of the input, up to a unitary transform \cite{jacobs.2002}. In this experiment, we use the entanglement generated by the biexciton radiative cascade, which results in the two-photon entangled state
\begin{equation}\label{ent}
{|\Psi(\tau)\rangle = \frac{1}{\sqrt{2}}\left[ |H_{XX}H_{X}\rangle + e^{i\Delta E \tau/\hbar} |V_{XX}V_{X}\rangle \right ]},
\end{equation}
where $H$ and $V$ denote the horizontal and vertical polarization eigenbasis of the QD. The telportation protocol is illustrated in the sketch of our experimental implementation in Fig. 3(a). While similar setups have been successfully used to teleport photons at shorter wavelengths\cite{Nilsson.2013, Huwer.2017}, no sources have so far been available to conduct this experiment in the technologically important wavelength region around 1550 nm. Here, the X photon is overlapped with the input laser qubit (Alice) at the 99:1 beamsplitter, with the intensity of the laser set to match the intensity of the QD to optimise both TPI performance and 3-photon coincidence count rates. We then perform a Bell-state measurement using a polarising beamsplitter (PBS) and detectors $D_{H}$ and $D_{V}$ calibrated to the QD eigenbasis (Charlie). The XX photon is finally analysed at a PBS by aligning the detectors $D_{P}$ and $D_{Q}$ to the expected output state basis (Bob). Experimentally, third-order correlations are measured at Charlie ($\tau_{Charlie}$) and Bob ($\tau_{Bob}$) both triggered from $D_{H}$. A successful Bell-state measurement heralds the telportation ($\tau_{Charlie}=0$). The teleportation fidelity is then calculated as $F_{P}=g^{(3)}_{P}/(g^{(3)}_{P}+g^{(3)}_{Q})$
where $g^{(3)}_{P(Q)}$ is the third-order correlation function for Bob's photon detected in polarization $P$($Q$), corresponding to the expected (unexpected) output polarization. By sending horizontal ($H$), vertical ($V$), diagonal $\left(D=(H+V)/\sqrt{2}\right)$, antidiagonal $\left(A=(H-V)/\sqrt{2}\right)$, right-hand circular $\left(R=(H-iV)/\sqrt{2}\right)$ and left-hand circular $\left(L=(H+iV)/\sqrt{2}\right)$ weak coherent laser input qubits, corresponding to the six symmetrically distributed polarization states on the Poincar\'{e} sphere, we can measure correlations at Bob to determine the mean fidelity of our teleporter.\\
For the dot presented in Fig. 2, driven at $P_{sat}$, the resulting mean fidelity coincidence map for the post-selection window corresponding to the most statistically significant teleportation is displayed in Fig. 3(b), where a fidelity of $83.6~\pm~2.2\%$ for an equivalent window size of 203 ps is achieved. This fidelity exceeds the classical threshold of 2/3 by 7.8 standard deviations, and is also 5.1 standard deviations above the limit imposed for secure implementations of 6-state secret key sharing protocols \cite{sixstate}.\\
\indent We use temporal post-selection to filter out the successfully teleported photons. Our teleporter is limited along $\tau_{Charlie}$ by the interference visibility of the laser and QD photon, with a timescale given by the $X$ coherence time and the detector resolution. Along $\tau_{Bob}$, the evolution of the quantum state due to the FSS, together with the detector resolution, limits the maximum window size. When varying the window size, smaller windows lead to higher fidelities at the cost of teleporting fewer photons within that window. For larger window sizes, more photons are teleported, but the signal starts to be washed out and the teleportation fidelity decreases.
Varying the temporal post-selection window size, we find the mean fidelity increases to a maximum of $88.4~\pm~4.0\%$ for an equivalent post-selection window size of 103 ps, which is 5.4 standard deviations above the classical threshold. This is the first reported value of quantum teleportation using a quantum dot emitting in this wavelength band. The point of highest fidelity in Fig. 3(b) is centered on $\tau_{Charlie}=\tau_{Bob}=0$ corresponding to the three-photon coincidence where the input polarization state is mapped to the output photon at Bob. Along the $\tau_{Charlie}=0$ axis we see oscillations in the fidelity resulting from the time-evolving nature of the two-photon entangled source. Indeed, the beat frequency is given exactly by the FSS of the QD. A cut-through the $\tau_{Bob}=0$ axis shows the high-fidelity teleportation heralding peak with width set by the X coherence time. It is clear in Fig. 3(b) a small amount of residual detuning of the laser from the QD is present, accounting for difference in the peak width from Fig. 2(d) in the TPI measurement. This is most likely as a result the 5.7 $\upmu$eV splitting combined with the $\sim2 \upmu$eV accuracy of the overlapping routine. Individual basis fidelities are displayed in Fig. 3(c), where each one is above both the classical and 6-state protocol thresholds. Furthermore, we note that due to the coherence time we are able to teleport superposition basis states with fidelity nearing the maximum given by the polar states, where TPI is not a requirement \cite{Nilsson.2013} and the teleportation fidelity is limited only by the entanglement fidelity. As such, our teleporter offers a robust and universal platform where teleportation of any arbitrary input state can be realised with high fidelity.

\section{Conclusion}
We have demonstrated the coherent emission from a telecom C-band quantum light source based on semiconductor QDs by measuring a single-photon coherence time exceeding 1~ns at low excitation powers. This compares well with state-of-the-art systems even when we use non-optimal, but highly practical, non-resonant excitation schemes. We have found that the source remains highly coherent even under power-broadened excitation and thus enables us to perform TPI measurements using a weak coherent source with near perfect visibility, when taking into account the limitations imposed by the multi-photon emission probability of the QD. We further showed how these experiments can be combined to form basic building blocks of a quantum network, by measuring proof-of-principle quantum teleportation of telecom C-band qubits.\\
\indent The teleportation shown here with emission wavelength in the telecom C-band straightforwardly enables extension of the reach of existing QKD systems, using simple non-resonant driving and relying only on standard industry growth and fabrication techniques in a material compatible with on-chip integrated detection. The low degree of spectral wandering and stable transition energy revealed by the long coherence times will further facilitate resonant excitation schemes, where we expect the true limits of coherence in emitted photons and the underlying spin states to be revealed. This will make our system available for a whole range of applications pioneered using short wavelength solid-state systems, at a wavelength that can be directly integrated with long-distance quantum networks.

\section{Acknowledgments}
The authors acknowledge partial financial support from the Engineering and Physical Sciences Research Council and the UK's innovation agency, Innovate UK. M. A. gratefully acknowledges support from the Industrial CASE award funded by the EPSRC and Toshiba Research Europe Limited.

%

\end{document}